\documentclass[11pt]{article}
\usepackage{amssymb,latexsym,amsmath,amsthm,graphicx,amscd}

\makeatletter \@addtoreset{figure}{section}
 \def\thefigure{\thesection.\@arabic\c@figure}
\def\fps@figure{h, t}
\@addtoreset{table}{bsection}
\def\thetable{\thesection.\@arabic\c@table}
\def\fps@table{h, t}
\newtheorem{theorem}{Theorem}

\newtheorem{proposition}[theorem]{Proposition}

\newfont{\tenbi}{cmbxti10}
\newcommand{\la}{\lambda}
\newcommand{\ga}{\gamma}
\newcommand{\eps}{\varepsilon}

\DeclareMathOperator{\Prym}{Prym}
\DeclareMathOperator{\Jac}{Jac}

\title{A new approach to separation of variables for the Clebsch integrable system. Part II: Inversion of the Abel--Prym map}
\author{ Y. Fedorov$^{1}$, F. Magri$^{2}$, T. Skrypnyk$^{3,4}$ \\
{\small $^{1}$ Polytechnic University of Catalonia, Barcelona, Spain}\\
{\small $^{2}$ Dipartimento di Matematica e Applicazioni- Universit\'{a} di Milano Bicocca, Milano, Italia}\\
{\small $^3$ Universit\'{a} degli Studi di Torino, via Carlo Alberto 10, 10123, Torino, Italia }\\
{\small $^{4}$ Bogolyubov Institute for Theoretical Physics, Metrolohichna str.14-b, 03115, Kiev, Ukraine}\\
{\footnotesize yuri.fedorov@upc.edu, franco.magri$@$unimib.it, taras.skrypnyk@unimib.it}}
\date{}

\begin{document}
\maketitle

\begin{abstract}
This is the second part of a paper describing a new concept of separation of variables applied to the classical Clebsch integrable case. The quadratures obtained in Part I lead to a new type of the Abel map which contains Abelian integrals on two different algebraic curves. 

Here we show that this map is from the product of the two curves
to the Prym variety of one of them, that it is well defined,
although not a bijection. We analyse its properties and formulate a new extention of the Riemann vanishing theorem, which allows to invert the map in terms of theta-functions of higher order. 

Lastly, we describe how to express the original variables of the Clebsch system in terms of the preimages of the map. This enables one to obtain theta-function solution for the system.      
\end{abstract}
\noindent


\section{Introduction: Abelian varieties and curves related to the quadratures.}

As was shown in Part I, to any point $(S_\alpha,T_\alpha)$, $\alpha=1,2,3$ of a generic invariant torus of the Clebsch system with the constants of motion $H_p, H_s, C_1, K_1$ there correspond 8 sets of separating variables $\{x_1,x_2\}$ and the corresponding conjugated momenta. Their   
 evolution with respect to the time parameters $t_s, t_p$ of the flows $X_s, X_p$ is described by the same quadratures, whose integral form reads (see formulas (9), (10) in Part I)
\begin{align}\label{Ab}
\begin{split}
\int_{x_{10}}^{x_1} \frac{d x}{w( w^2 - g(x) ) }+ 
\int_{x_{20}}^{x_2} \frac{\sqrt{2} \, dx}{W (W^2 - g(x))} &= \frac{i}{ C_1} t_{s} \, ,\\
\int_{x_{10}}^{x_1}\frac{x \,d x }{w( w^2 - g(x) )  }+ 
\int_{x_{20}}^{x_2} \frac{\sqrt{2}\, x\, d x}{W (W^2 - g(x) ) }&=\frac{i}{ C_1} t_{p}\, , \qquad 
i =\sqrt{-1}\, ,
\end{split}
\end{align}
where $g(x) = K_1 x^2+ H_p x + H_s$ and $x_{10}, x_{20}$ are some initial values. (In both Parts we assume that $C_1 \ne 0$.)   

Here the pairs $(x_1,w)$ and $(x_2,W)$ satisfy equations of algebraic curves 
\begin{align}
\begin{split}
 &C : \qquad w^4 - 2 g(x_1)w^2 + g^2(x_1) - 4 C_1^2 (x_1+j_1)(x_1+j_2)(x_1+j_3) = 0 \,, \\ \label{curves}
 &K : \qquad W^4 -2 g(x_2) W^2 + 4 C_1^2 (x_2+j_1)(x_2+j_2)(x_2+j_3) =0\, . 
 \end{split}
\end{align}
As one can check, both curves are non-hyperelliptic of genus 3, and they are not birationally equivalent. The left hand sides of \eqref{Ab} involve two holomorphic differentials on the curve $C$ and two holomorphic differentials on $K$.   

The form of the quadratures reminds the standard Abel--Jacobi map $\Gamma^{(2)} \to \Jac(\Gamma)$, where $\Gamma^{(2)}$ is the symmetric product of two copies of a regular genus 2 curve $\Gamma$ and $\Jac(\Gamma)$ is the common notation for the Jacobian variety of $\Gamma$ (see, e.g., \cite{GH, MumTh}). Such a map can be inverted: in particular, any symmetric function of coordinates of the two points on $\Gamma$ can be expressed in terms of the Riemann theta-functions of $\Gamma$ (see, e.g., \cite{Dub}).  

However, the facts that in the quadratures \eqref{Ab} the periods of the Abelian integrals on the two curves are distinct and the genus of $C$ and $K$ is higher than the dimension of the invariant tori (and the number of equations in \eqref{Ab}) raise a natural doubt about invertibility of these quadratures in terms of meromorphic functions of the complex times $t_s, t_p$. 

On the other hand, it is known (\cite{avm84}) that the Clebsch integrable case is an algebraic integrable system, which implies that all its solutions are meromorphic.

In Section 1 of this Part II we will show that the periods of the integrals in \eqref{Ab}, although being different, are commensurable, and that the quadratures lead to a well-defined, so called Abel--Prym map from the product $C\times K$ to a two-dimensional Prym subvariety of the Jacobian of one of the curves.  

In Section 2 we introduce higher order theta-functions and formulate analogs of the Riemann vanishing theorem, which will be a main tool of the inversion of the map. We will see that, in contrast to the standard Abel map, the Abel--Prym maps are not one-to-one: a full preimage of a point in the Prym variety consists of 8 pairs of points on $C\times K$.   
 
In Section 3 we prove that the coordinates of these 8 pairs of points can be identified with the eight sets of separating variables for the Clebsch system constructed in Sections 3,4 of Part I. This enables one to derive a (new) theta-function solution for the original variables $S_\alpha, T_\alpha$: they can be expressed as symmetric functions of coordinates of the eight points on one of the curves, while the latter functions can be written in terms of the higher order theta-functions which solve the problem of inversion of the Abel--Prym map. 
(We will only give a sketch of this procedure, as explicit expressions are too long to present.)    
\medskip

We start with recalling necessary properties
of the genus 3 curves $C$ and $K$ and their relation to the complex invariant tori of the Clebsch system, already described in \cite{hai83, ef17}\footnote{The analysis in \cite{hai83} has been made for other pair of genus 3 curves related to the integrable Frahm--Manakov top on $so(4)$, however both pairs of curves are birationally equivament.}

First, observe that the curves $K$ and $C$ actually already appeared as the spectral curves of two different Lax representations of the Clebsch system obtained in \cite{Per}
and, respectively, in \cite{Bob_so4} (see also \cite{Bel,Fe_AMS}). Equations \eqref{curves} can be written in the form
\begin{align}  
C &: \;\;  w^2 = g(x) + 2 C_1 \sqrt{\Phi(x)}, \qquad
 \Phi =(x+j_1)(x+j_2)(x+j_3),  \label{mastercurve} \\
K &: \; W^2 = g(x) + \sqrt{\Psi(x)} , \label{dual_K}
\\
\Psi (x) & =g^2(x)-4 C_1^2 \Phi(x) = K_1^2(x-s_{1}) (x-s_{2})(x-s_{3})(x-s_{4}) ,   \notag
\end{align}
which makes evident that the curves $C,K$ have the involution $\sigma \, : (x_1,w)\to (x_1,-w)$, respectively, $\sigma \, : (x_2,W)\to (x_2,-W)$, and they are 2-fold ramified coverings of elliptic curves $E, {\cal E}$: 
$$
\pi\,: \; C \to E= \{y^2= \Phi(x_1)\}, \quad
\bar\pi\,:\; K\to {\cal E} =  \{y^2= \Psi(x_2)\}.
$$
The coverings are ramified at points $Q_1,\dots Q_4\in E$, respectively $Z_1,\dots, Z_4 \in {\cal E}$, so that
$\sigma(Q_j)=Q_j, \sigma(Z_j)=Z_j$ (one of $Z_j$ is an infinite point of $K$). Note that in our case $C_1 \ne 0$
these points are not in the hyperelliptic involution $(x,y)\to (x,-y)$ acting on $E$, respectively on $\cal E$.

The involution $\sigma : (x,w)\to (x,-w)$ on $C$ extends to its Jacobian variety,
$\Jac (C)$. Thus the latter contains two Abelian subvarieties: the elliptic curve $E$ itself and the 2-dimensional
Prym variety denoted as $\Prym(C,\sigma)$, which is anti-symmetric with respect to the extended involution, whereas
$E$ is invariant. Equivalently, \\ $\Prym(C,\sigma)=\text{ker}\, (1+\sigma)$, see e.g., \cite{Mum}.
One should stress that $\Prym(C,\sigma)$ is not the Jacobian variety of a genus 2 curves because it has polarization $(1,2)$.

Similarly, one defines the Prym variety $\Prym(K,\sigma)\subset \Jac(K)$,
which is anti-symmetric with respect to the  involution $\sigma$ extended to the Jacobian.

\begin{theorem}[L. Haine \cite{hai83}] \label{Th1} \begin{description}
\item{1)} The complex invariant manifold ${\cal I}_H$ of the Clebsch system with generic constants of motion $H_p,H_s,C_1,K_1$ is isomorphic to an open subset of $\Prym(C,\sigma)$. Namely, ${\cal I}_H = \Prym(C,\sigma)\setminus {\cal D}$, where
${\cal D}$ is a genus 9 curve wrapped around the complex torus $\Prym(C,\sigma)$.
The variables $S_\alpha, T_\alpha$ are meromorphic functions on $\Prym(C,\sigma)$ having a simple pole along $\cal D$.

\item{2)}There is a 4-fold unramified covering (an isogeny)
$\Pi^*\, : \; \Prym(C,\sigma) \to \Prym(K,\sigma)$, which is associated with the action of discrete group $\mathfrak g$ of 4 elements on $\Prym(C,\sigma)$, which change signs of some of $S_\alpha, T_\alpha$. Then
$\Prym(C,\sigma)/\mathfrak g =\Prym(K,\sigma)$. The curve
$\cal D$ can be viewed as 4-fold unramified covering
$\Pi\,: {\cal D}\to C={\cal D}/{\mathfrak g}$.

Thus the squares $S_\alpha^2, T_\alpha^2$ are meromorphic functions on $\Prym(K,\sigma)$, having a second order pole along $C\subset \Prym(K,\sigma)$.

\item{3)} Let $\widehat E$ be the 4-fold unramified covering of the elliptic curve $E=C/\sigma$, obtained by doubling its two periods. Then $\cal D$ can be viewed as 2-fold covering of $\widehat E$ ramified at the 16 preimages of
    the branch points $Q_1,\dots, Q_4\in E$, as described in the diagram
$$
\begin{CD}
\Pi\,:\; {\cal D} @ > 4:1 \; \text{ur} >>  C \\
       @  V 2:1 \; \text{ram. at 16 $\Pi^{-1}(Q_i)$} VV  @  V  VV 2:1  \; \text{ram. at 4 $Q_i$} \\
  \widehat E @ > 4:1 \; \text{ur} >>  E=C/\sigma
 \end{CD}
$$
where {\tt ur} and {\tt ram} mean ''unramified'' and ''ramified'' respectively.

\item{4)} The complex time flows $X_s, X_p$ of the Clebsch system are straight lines on $\Prym(C,\sigma)$.
\end{description}
\end{theorem}

The curve $\cal D$ will play an important role in the sequel. 
As a 4-fold covering of $C$, it can be described as a spacial curve in ${\mathbb C}^5(x,w,v_1,v_2,v_3)$ given by the conditions 
\begin{equation} \label{D1}
{\cal D}\, : \;  w^2 = g(x) + 2 C_1 \,v_1 v_2 v_3, \quad v_1^2 =x+j_1, \quad
v_2^2=x+j_2, \quad v_3^2 =x+j_3.
\end{equation}
They are obtained from the equation \eqref{mastercurve} of $C$. 
Indeed, any point $(x,w)$ on $C$ specifies a value of the product $v_1 v_2 v_3$, leaving 4 choices of signs of $v_1,v_2,v_3$.  The projection $\Pi\, :\; {\cal D} \to C$ simply reads as:  $(x,w,v_1,v_2,v_3) \to (x,w)$. 
 
Alternatively, substituting in \eqref{D1} the expressions (24, Part I) for $v_i, x$ in terms of the coordinates on the big elliptic curve $\hat E=\{ Y^2= 4 (Z+j_1)(Z+j_2)(Z+j_3)\}$  (23, Part I), namely
\begin{align}
  v_\alpha &= \frac{Z^2+2 j_\alpha Z + j_\alpha(j_\beta+j_\gamma)- j_\beta j_\gamma}{Y}, \qquad
(\alpha,\beta,\gamma)=(1,2,3), \label{vi} \\
 x &= \frac{(Z^2 -j_1 j_2-j_2 j_3-j_3 j_1)^2-4 j_1 j_2j_3
(j_1+j_2+j_3)}{ Y^2} , \notag
\end{align}
one obtains the equations defining $\cal D$ in ${\mathbb C}^3(Z,Y,W)$; it is now viewed as the 2-fold ramified covering of $\hat E$:
\begin{gather}  \label{D2}
 G^2 =p_7(Z)+ 2 C_1\,Y\, p_6(Z), \quad
Y^2 = 4(Z+j_1)(Z+j_2)(Z+j_3),
\end{gather}
where $G= w Y^2$ and $p_6(Z), p_7(Z)$ are certain polynomials of degrees 4 and 6.

The above results can be completed with the following theorem.

\begin{theorem}[W. Barth, \cite{bw85}] \begin{description}
\item{1)} $\Prym(K,\sigma) \subset \Jac(K)$ contains the curve $C$ as the intersection $\Prym(K,\sigma) \cap \Theta_K$, where $\Theta_K \subset \Jac(K)$ is a translate of the theta-divisor of curve $K$.
Moreover,  $C \subset \Prym(K,\sigma)$ belongs to the pencil $\{C_\lambda\}$, $\la \in {\mathbb P}$ of curves of genus $\le 3$ given by intersections $\Prym(K,\sigma) \cap \Theta_{K, \la}$, where $\Theta_{K, \la}$ are translations of $\Theta_K$ along the elliptic curve ${\cal E} \subset \Jac(K)$. The base points of $\{C_\lambda\}$ are precisely the branch points $Q_1,\dots, Q_4$ of the covering $C\to E$.

\item{2)} For any curve $C_\la\subset\Prym(K,\sigma)$, the involution
$\sigma: \,C_\la \to C_\la$ has the same fixed points $Q_1,\dots, Q_4$. Thus a generic curve $C_\la$ is a 2-fold covering
of an elliptic curve $E_\la$, and the corresponding varieties $\Prym(C_\la,\sigma)$ are all isomorphic.

\item{3)} Similarly, $\Prym(C,\sigma) \subset \Jac(C)$ contains
the curve $K$ as the intersection $\Prym(C,\sigma) \cap \Theta_{C}$;  $K$ belongs to the pencil $\{K_\lambda\}$, $\la \in {\mathbb P}$ of curves of genus $\le 3$ as intersections $\Prym(C,\sigma) \cap \Theta_{C, \la}$, where $\Theta_{C, \la}$ are translations of the
theta divisor $\Theta_C$ along the elliptic curve $E \subset \Jac(C)$.
The base points of $\{K_\lambda\}$ are the branch points $Z_1,\dots, Z_4$ of the covering $K\to {\cal E}$.
\end{description}
\end{theorem}

\paragraph{The periods of the Prym varieties and the Abel--Prym map.}
In accordance to the equations \eqref{Ab}, choose the following basis of the holomorphic differentials on the curves $C$ and $K$:
\begin{gather}
 \omega_1=  \frac{d\, x}{2 C_1\, w \sqrt{\Phi(x)} } =\frac{d\, x}{w(w^2-g(x))} , \quad
\omega_2= \frac{x \,d\, x}{2 C_1\, w \sqrt{\Phi(x)} } =\frac{x\, d\, x}{w(w^2-g(x))}, \label{WW} \\
\omega_0=\frac{d\,x}{2\sqrt{\Phi(x)}},\notag
\end{gather}
and, respectively,
\begin{gather}
 \bar \omega_1 = \frac{\sqrt{2}\, d\, x}{W \sqrt{\Psi(x)} } =
 \frac{\sqrt{2}\, d\, x}{W (W^2-g(x))}, \quad
\bar \omega_2=  \frac{\sqrt{2}\, x \,d\, x}{W \sqrt{\Psi(x)} }
= \frac{\sqrt{2}\, x \,d\, x}{W (W^2-g(x))},  \label{ww} \\
\bar\omega_0=\frac{d\,x}{\sqrt{\Psi(x)} } \, .\notag
\end{gather}
The differentials $\omega_1, \omega_2, \bar \omega_1, \bar \omega_2$ are anti-symmetric with respect to involution $\sigma$, whereas
$\omega_0, \bar\omega_0$ are symmetric.

Let cycles $\ga_1, \dots \ga_6$ and $\bar\ga_1, \dots \bar\ga_6$
form canonical bases of cycles in $H_1(C,{\mathbb Z})$ and, respectively, in $H_1(K,{\mathbb Z})$.

\begin{theorem} \label{2periods} \begin{description}
\item{1)} The 6 period vectors
$V_i= \oint_{\ga_i} \begin{pmatrix} \omega_1 \\ \omega_2 \end{pmatrix} \in {\mathbb C}^2(u_1, u_2), \; i=1,\dots,6$ form a lattice $\Lambda$ of rank 4.
The complex torus ${\mathbb C}^2/\Lambda$ is isomorphic to the Prym variety $\Prym(K,\sigma)$.

In the same space ${\mathbb C}^2(u_1, u_2)$, the 6 period vectors
$\bar V_i= \oint_{\bar\ga_i} \begin{pmatrix} \bar \omega_1 \\ \bar \omega_2 \end{pmatrix}, \; i=1,\dots,6$ form a lattice $\bar\Lambda$ of rank 4, and
$\Prym(C,\sigma)= {\mathbb C}^2/\bar\Lambda$.

Further, chose $\{\ga_1,\dots, \ga_6\}=\{a,A,b,B,\bar a, \bar b\}$ such that the cycles $(a,b), (\bar a,\bar b)$ and $A,B$ are pairwise conjugated and
$$
\sigma(A)=-A,\quad \sigma(B)=-B, \quad \sigma(a)= \bar a, \quad \sigma(b)= \bar b,
$$
then for the basis $\{\ga_1,\dots, \ga_4\}=\{a,A,b,B\}$ the period matrix of $\Prym(K,\sigma)$ reads
$ \Omega= (V_1\, V_2\, V_3 \, V_4)$.

For the similar choice of canonically conjugated cycles $\bar\ga_1,\dots, \bar \ga_4$ on $K$, the period matrix of
$\Prym(C,\sigma)$ is
$\bar\Omega = (\bar V_1\, \bar V_2\, \bar V_3 \, \bar V_4)$.

\item{2)} The above period matrices are related as follows
\begin{equation} \label{latt}
\bar\Omega= (\bar V_1\, \bar V_2\, \bar V_3 \, \bar V_4)= (2V_1\; V_2\; 2 V_3 \; V_4).
\end{equation}
\end{description}
\end{theorem}

Thus, the periods of the holomorphic differentials \eqref{WW}, \eqref{ww} on $K,C$ are conmensurable, which implies that the quadratures \eqref{Ab} can be inverted in terms of meromorphic functions.

Note that the relation \eqref{latt} is consistent with Theorem \ref{Th1} saying the torus $\Prym(C,\sigma)$ is a 4-fold unramified covering of $\Prym(K,\sigma)$. One should stress however that \eqref{latt} holds only if anti-symmetric differentials on $C,K$ are both proportional to \eqref{WW}, \eqref{ww}.

\medskip
\noindent{\it Proof of Theorem} \ref{2periods}.
Item 1) was proven in \cite{hai83}, where it was shown that  $\Prym(K,\sigma)$ and $\Prym(C,\sigma)$ are dual Prym subvarieties.

Item 2) is a recent observation. Since $\Prym(C,\sigma)$ can be regarded as 4-fold covering of $\Prym(K,\sigma)$ (see Theorem 1), the relation \eqref{latt} between the period matrices must hold for certain bases of holomorphic anti-symmetric differentials on $K,C$. To prove that it holds for the bases
\eqref{WW}, \eqref{ww}, we use the geometric fact $K= \Prym(C,\sigma)\cap \Theta_C$ (item 3 of Theorem 2). Namely, let, as above $u_1, u_2, u_3$ be the coordinates in ${\mathbb C}^3$, the universal covering of $\Jac(C)$, and
$du_1, du_2$ be the corresponding holomorphic $\sigma$-anti-symmetric one-forms on $\Prym(C,\sigma) \subset \Jac(C)$ such that
$$
 \omega_1 = du_1 \big |_C, \quad  \omega_2 = du_2 \big |_C\, .
$$
Then relation \eqref{latt} is geometrically equivalent to
\begin{equation} \label{red_diffs}
du_1 \big |_K =\bar\omega_1, \quad du_2 \big |_K=\bar  \omega_2\, .
\end{equation}

To prove \eqref{red_diffs} explicitly, we use the following algebraic description of the theta-divisor
$$
 \Theta_C = \{ P_1+P_2-\infty_1-\infty_2 \mid P_i=(x_i,w_i)\in C\},
$$
$\infty_{1,2}$ being the two points on $C$ with $x=\infty$, $\sigma(\infty_1)=\infty_2$,
and of the embedding $K \hookrightarrow \Theta_C$, which was described in \cite{Pant} and made explicit in Appendix 5 of \cite{au96}:
\begin{equation} \label{embedding_K}
K \ni (x,W) \mapsto D= (x_1,w_1)+ (x_2,w_2)- \infty_1 - \infty_2 \in \Theta_C,
\end{equation}
where now $x_1=x_2=x$ and $w_1,w_2$ are the solutions of the quadratic equation
\begin{equation} \label{r(w)}
r(w)= w^2-W \sqrt{2} w+ W^2-g(x)=0.
\end{equation}
As one can check, the points $(x,w_1), (x,w_2)$ belong to $C$, and $\{w_1, w_2, -w_1,-w_2\}$ give all the solutions 
of the equation of $C$ in \eqref{curves} for a given $x$. Then
$$
 D + \sigma(D) = (x,w_1)+ (x,w_2)+ (x,-w_1)+ (x,-w_2) -   2\infty_1 - 2 \infty_2
$$
is the divisor of a meromorphic function $f$ on $C$ having 4 zeros over $x$ and double poles at $\infty_1, \infty_2$, that is, $D+ \sigma(D)\equiv 0$. Thus for any point of $K$ its image $D$ belongs both to
$\Prym(C,\sigma)$ and to $\Theta_C$. One can also show that the map
\eqref{embedding_K} is injective and surjective, hence it describes the isomorphism between $K$ and $\Prym(C,\sigma)\cap\Theta_C$.

By the definition of $du_1, du_2$, and of the theta-divisor, on $\Theta_C \subset \Jac(C)$ one has
$$
\begin{aligned}
  du_1 \big |_{\Theta_C} & = \frac{dx_1}{w_1 (w_1^2-g(x_1))} +
\frac{dx_2}{w_2 (w_2^2-g(x_2)) } , \\
 du_2 \big |_{\Theta_C} & = \frac{x_1\, dx_1}{w_1 (w_1^2-g(x_1))} + \frac{x_1\,dx_2}{w_2 (w_2^2-g(x_2)) } .
\end{aligned}
$$
On $K\subset \Theta_C$ we have $x_1=x_2$ and
\begin{equation} \label{du12}
du_1 |_K =
\frac{ w_1^3 + w_2^3 -g(x)(w_1+ w_2)}{w_1 w_2 (w_2^2-g(x)) (w_2^2-g(x))} \,dx   \quad
du_2 |_K =
\frac{ w_1^3 + w_2^3 -g(x)(w_1+ w_2)}{w_1 w_2 (w_2^2-g(x)) (w_2^2-g(x))} x \,dx \,, 
\end{equation}
where now the coordinates
$w_1,w_2$ are the roots of the quadratic equation \eqref{r(w)}, hence their symmetric functions become functions of $x,W$. Then, after simplifications, the right hand sides of \eqref{du12} yield
$$
du_1 |_K = \sqrt{2} \frac{dx}{W (W^2-g(x))}=\bar \omega_1, \quad
du_2 |_K =\sqrt{2} \frac{x\, dx}{W (W^2-g(x))}=\bar \omega_2,
$$
which proves \eqref{red_diffs} and item (2) of the theorem. $\square$
\medskip

\paragraph{The Abel--Prym map.}
Analytically, the curve $C \subset \Prym(K,\sigma)$ can be viewed as the image of the smooth
embedding\footnote{Note that the differentials $\omega_1, \omega_2$ in \eqref{ww} do not have common zeros on $C$.}
\begin{equation} \label{calA}
  {\cal A}\, : \; P \in C \mapsto   \int_{P_0}^P   \begin{pmatrix} \omega_1 \\ \omega_2 \end{pmatrix} \quad \text{mod $\Lambda$}\,
\end{equation}
where $P_0\in C$ is any fixed basepoint and $\Lambda$ is the period lattice described in Theorem 3. For concreteness, we choose $P_0$ to be one of the branch points of the covering $\pi:\, C\to E$, say $P=Q_1$.
Then $\cal A$ sends $Q_1$ to the origin, and the points $Q_2,Q_3, Q_4$ to some half-periods of  $\Prym(K,\sigma)$.


By analogy with $\cal A$, we also define the map
\begin{align} \label{barA}
\bar {\cal A}\,: \; K & \mapsto \Prym(C,\sigma)={\mathbb C}^2/\bar\Lambda, \\
 R \in K & \mapsto \int_{Z_1}^R \begin{pmatrix} \bar \omega_1 \\ \bar \omega_2 \end{pmatrix} \quad \text{mod $\bar\Lambda$} \notag
\end{align}
giving a smooth embedding of $K$ to $\Prym(C,\sigma)$.
The image $\hat K= \Pi\circ \bar {\cal A} (K)\subset \Prym(K,\sigma)$ is a curve with self-intersections.

Then the  quadratures \eqref{Ab} for the Clebsch system give rise to the map
\begin{align} \label{APM}
{\cal P}\, : \; C\times K &\to \Prym(K,\sigma) \\
  P \in C,\; R \in K &\mapsto  \int_{Q_1}^P  \begin{pmatrix} \omega_1 \\ \omega_2 \end{pmatrix} +
 \int_{Z_1}^R  \begin{pmatrix} \bar \omega_1 \\ \bar \omega_2 \end{pmatrix} = \begin{pmatrix} u_1 \\ u_2 \end{pmatrix}
\quad \text{mod $\Lambda$}. \notag
\end{align}

According to Theorem \ref{2periods}, $\cal P$ is
well-defined:
when the point $P$ goes along a full cycle on $C$, the first integral in \eqref{APM} changes by a vector of
the lattice $\Lambda$; and when the point $R$ goes along a full cycle on $K$, the second integral in \eqref{APM} changes by a vector of $\bar\Lambda$, which is a sub-lattice of $\Lambda$.

In the sequel it is natural to call $\cal P$ the {\it Abel--Prym} map\footnote{Not to be confused with the {\it Prym map},
which has a completely different meaning in algebraic geometry.}. As we shall see below, in contrast to the standard
Abel--Jacobi map, $\cal P$ is not injective, so, for a generic ${\bf u}=(u_1, u_2)\in \Prym(K,\sigma)$, its inversion
is not unique. To our best knowledge, such kind of map did not appear  before, neither in the classical nor in the modern
literature.

\paragraph{The extended Abel--Prym map.}
The map $\cal P$ cannot be extended to the map \\ $C\times K \to \Prym(C,\sigma)$ as the latter is not well-defined: under the map $\cal A$, one and the same point in $C$ yields 4 different points in $\Prym(C,\sigma)$. In this connection it is natural to replace the curve $C$ by its 4-fold covering $\cal D$ described in Theorem 1 and \eqref{D1}, \eqref{D2}, and to introduce the map $\widehat{\cal A}\,:\, {\cal D} \mapsto \Prym(C,\sigma)$:
$$
   P\in {\cal D} \to \widehat{\cal A} (P) = \int_{Q_1}^{\Pi(P)} \begin{pmatrix}  \omega_1 \\ \omega_2 \end{pmatrix}
\qquad mod \; \bar\Lambda,
$$
where, as above, $\Pi$ is the projection ${\cal D}\to C$.

 Note that for a general genus 9 curve,
the standard Abel map requires integrals of 9 holomorphic differentials on it, otherwise it is not
correctly defined. However, as was shown in \cite{hai83}, in case of $\cal D$ the following property holds.

\begin{proposition}
The map $\widehat{\cal A}$ is injective and realizes a smooth isomorphism between $\cal D$ and its image
$\widehat{\cal A}({\cal D})\subset \Prym(C,\sigma)$. It can also be written as
 $$
   P\in {\cal D} \to \widehat{\cal A} (P) = \int_{Q_1^*}^{P}  \begin{pmatrix} \Pi^*(\omega_1) \\ \Pi^*(\omega_2) \end{pmatrix}
\qquad \textup{mod }\; \bar\Lambda,
$$
where $\Pi^*(\omega_j), \, j=1,2$ are the pull-backs of the holomorphic differentials $\omega_j$ on $C$, and
$Q_1^*$ is one of the preimages $\Pi^{-1}(Q_1)$.
\end{proposition}

Now introduce {\it extended Abel--Prym map} $\widehat{\cal P}\, : \, {\cal D}\times K \to \Prym(C,\sigma)$:
\begin{equation} \label{eAPm}
\widehat{\cal P}\, :\; (P\in {\cal D}, R\in K) \mapsto {\bf u} =(u_1, u_2)^t=\widehat {\cal A}(P)+ \bar {\cal A}(R) ,
\end{equation}
$\bar {\cal A}$ being defined in \eqref{barA}.
Like $\cal P$, the above map $\widehat{\cal P}$ is
well-defined: 
when $P$ goes along a full cycle on $\cal D$ or when the point $R$ goes along a full cycle on $K$,
the integrals in \eqref{eAPm} change by a vector of the lattice $\bar\Lambda$.

Note (without a proof) that, for any $(P,R) \in  {\cal D}\times K $,
\begin{equation} \label{comm}
\Pi^* \circ \widehat{\cal P} (P,R) = {\cal P} (\Pi(P),R) .
\end{equation}

\section{Inversion of the maps $\cal P$, $\hat{\cal P}$ by means of theta-functions}

We first build a set of theta-functions which will be used to invert the maps $\cal P$, $\widehat{\cal P}$.

As follows from item 2) of Theorem \ref{2periods}, 
there exists a unique change of variables $(u_1, u_2)\to (z_1, z_2)$, described by a
non-degenerate matrix $T\in GL(2,{\mathbb C})$, which takes the period matrices of $\Prym(K,\sigma)$, $\Prym(C,\sigma)$ to the normalized form
\begin{equation} \label{periods}
  T \cdot (V_1\, V_2\, V_3 \, V_4)
= \begin{pmatrix}  1 & 0 & a & b \\
                   0 & 2 & b & c \end{pmatrix}, \quad
 T \cdot (\bar V_1\, \bar V_2\, \bar V_3 \, \bar V_4)
= \begin{pmatrix}  2 & 0 & 2a & b \\
                   0 & 2 & 2b & c \end{pmatrix},
\end{equation}
with some $a,b,c$ satisfying the Riemann conditions,
and by the extra change $(z_1,z_2) \to (z_1,z_2/2)$, the latter matrix is transformed to the canonical form
$\begin{pmatrix}  2 & 0 & 2a & b \\
                   0 & 1 & b & c/2 \end{pmatrix}$.

As was shown in \cite{hvm89, ef17}, $\Prym(K,\sigma)$ is a 2-fold unramified covering of 3 different
principally polarized Abelian tori. As follows from the first period matrix in \eqref{periods},
one of them is the Jacobian of a genus 2 curve $\Gamma$ with the Riemann matrix
\begin{equation} \label{tau_m}
\tau =   \begin{pmatrix}   a & b \\ b & c \end{pmatrix}\, .
\end{equation}
Similarly, $\Prym(C,\sigma)$ is a 2-fold unramified covering of 3 other
principally polarized Abelian tori, and, following the second relation in \eqref{periods},
one of them is the Jacobian of a genus 2 curve $\tilde\Gamma$ with the period matrix
\begin{equation} \label{pers_Jacs}
 \begin{pmatrix}  2 & 0 & a & b \\
                   0 & 2 & b & c \end{pmatrix}, \; \text{giving, by the rescaling ${\bf z} \to 2 {\bf z}$, the matrix} \;
 \begin{pmatrix}  1 & 0 & a/2 & b/2 \\
                   0 & 1 & b/2 & c/2 \end{pmatrix}.
\end{equation}
Thus, the Riemann matrix of $\tilde\Gamma$ is $\tau/2$, and
one has the following chain of 2-fold coverings
\begin{equation} \label{puzzle}
\begin{CD}
\Prym(C,\sigma) @ > 2:1 >> \Jac(\tilde\Gamma)  @ > 2:1 >> \Prym(K,\sigma) @ > 2:1 >>  \Jac(\Gamma)  \, .
\end{CD}
\end{equation}

Explicit algebraic equation of $\Gamma$ in terms of the coefficients of $C$
was given by F. K\"otter in \cite{kot892} (see also \cite{ef17}), who linearized the Clebsch system on $\Jac(\Gamma)$.
In the sequel we will need only the matrix $\tau$, and not expressions for $\Gamma, \tilde \Gamma$ themselves.

Now consider the standard Riemann theta-function $\theta ({\bf z}\mid \tau )$,  ${\bf z}=(z_1, z_2)^t = T (u_1, u_2)^t$,
associated with the Riemann period matrix $\tau$ in \eqref{tau_m},
$$
\theta ({\bf z} \mid \tau )= \sum_{N\in {\mathbb Z}^2 } \exp\left( \pi\jmath \langle N ,\tau N\rangle
+2\pi \jmath \langle N, {\bf z}\rangle  \right), \quad \jmath=\sqrt{-1}
$$
and theta-functions with half-integer characteristics
 $ {\bf \alpha }=(\alpha_1, \alpha_2),  {\bf \beta }=(\beta_1, \beta_2) \in \frac 12 {\mathbb Z}^2/{\mathbb Z}^2$ defined by
$$
    \theta\! \left[ { \alpha \atop \beta } \right ]  ({\bf z} \mid \tau )
= \exp \left( \jmath \pi (\alpha \tau \alpha^t +2\beta \alpha^t+2\beta {\bf z} ) \right)
\theta ({\bf z} + \beta^t + \tau \alpha^t \mid \tau ) .
$$
Thus  $\theta\! \left[ { \alpha \atop \beta } \right ] \! ({\bf z} \mid \tau )$ is the Riemann theta-function with
the argument translated by the half-period $\beta^t + \tau \alpha^t$ and multiplied by an exponent. These functions
 have the quasiperiodic property: for any $N,M\in {\mathbb Z}^2, $ 
\begin{gather}
\theta\! \left[ { \alpha \atop \beta } \right ] \! ({\bf z}+ K+\tau M)
=\exp (2\pi \jmath\epsilon) \exp \{\langle M,\tau M\rangle/2+ \langle M,{\bf z}\rangle \}\,
\theta\! \left[ { \alpha \atop \beta } \right ]  \!({\bf z}) , \label{1.5} \\
\epsilon = \langle \alpha, K\rangle - \langle\beta, M\rangle , \nonumber
\end{gather}

It is known (\cite{MumTh,Bel,fay}) that the condition $\theta ({\bf z}\mid \tau )=0$ defines the theta-divisor $\Theta\subset \Jac(\Gamma)$ isomorphic to the curve $\Gamma$ itself, and $\theta ({\bf z}\mid \tau )$ vanishes at
6 odd half-periods in $\Jac(\Gamma)$ corresponding to the characteristics
 \begin{gather}
\Delta_0 =\begin{pmatrix} 1/2&1/2\cr0&1/2 \end{pmatrix} ,\quad
 \Delta_{1}=\begin{pmatrix} 0&1/2\cr0&1/2 \end{pmatrix}, \quad
\Delta_{2}=\begin{pmatrix} 0&1/2\cr1/2&1/2 \end{pmatrix}, \notag \\
\Delta_{3}=\begin{pmatrix} 1/2&0\cr1/2&1/2\end{pmatrix}, \quad
\Delta_{4}=\begin{pmatrix} 1/2&0\cr1/2&0\end{pmatrix}, \quad
 \Delta_{5}=\begin{pmatrix} 1/2&1/2\cr1/2&0 \end{pmatrix} .
\label{deltas}
\end{gather}

\paragraph{The quasi-periodic theta-functions on $\Prym(K,\sigma)$.} Let us set
\begin{equation} \label{tts}
 \theta_0 ({\bf z}) = \theta  \left[\begin{matrix} 0 & 0 \\ 0 & 0 \end{matrix} \right] ({\bf z}+ {\cal K} \mid \tau), \quad
\theta_1 ({\bf z}) = \theta  \left[\begin{matrix} 0 & 1/2 \\ 0 & 0 \end{matrix} \right] ({\bf z}+ {\cal K} \mid \tau),
\end{equation}
where ${\cal K}= (1/2,0)^t+ \tau (1/2,1/2)^t$ is the vector of Riemann constants, the half-period corresponding to the characteristic $\Delta_0$ (see, e.g., \cite{MumTh}).

 In view of \eqref{1.5}, for a generic quotient $(\nu: \mu)\in {\mathbb P}$, the function
$$
  \Xi_{\nu, \mu} ({\bf z}) = \nu  \theta_0 ({\bf z}) +  \mu \theta_1 ({\bf z})
$$
is not quasiperiodic on $\Jac(\Gamma)$, but on its 2-fold covering $ {\mathbb T}^2 \to \Jac(\Gamma)$, obtained by doubling the period vector $(0,1)^t$ of  $\Jac(\Gamma)$.
Indeed,
$$
 \theta_0 \left({\bf z}+ (0,1)^t \right)
=\theta_0 ({\bf z}), \quad  \theta_1 \left({\bf z}+ (0,1)^t \right) =- \theta_1 ({\bf z}),
$$
and if $\bf z$ changes by any other period vector, $ \theta_0, \theta_1$ are multiplied by the same factor.
In view of the first matrix in \eqref{pers_Jacs}, ${\mathbb T}^2$ is precisely $\Prym(K,\sigma)$.
Hence the zeros of  $\Xi_{\nu, \mu} ({\bf z})$ on
 $\Prym(K,\sigma)$ are well-defined. Following \cite{hvm89},
equation $\Xi_{\nu, \mu} ({\bf z})=0$ defines the corresponding curve of the pencil $\{C_\la| \la=\mu/\nu\in {\mathbb C}\}$ described in Theorem 2.

Note that $\Xi_{\nu, \mu} ({\bf z})$ can be regarded as a higher-order theta-function (or a Prym theta-function) introduced in \cite{fay}.

Since we changed to the normalized coordinates $(z_1, z_2)^t = T (u_1, u_2)^t$ on the Pryms, we now redefine the maps \eqref{calA}, \eqref{barA} accordingly:
$$
{\cal A}\, : \; P \in C \mapsto   \int_{Q_1}^P   T \begin{pmatrix} \omega_1 \\ \omega_2 \end{pmatrix}, \quad
\bar {\cal A}\, : \; R \in K \mapsto   \int_{Z_1}^R T \begin{pmatrix} \bar \omega_1 \\ \bar \omega_2 \end{pmatrix} .
$$

Observe that the images $ {\cal A}(Q_1), \dots , {\cal A}(Q_4)\in \Prym(K,\sigma)$ of the basepoints
$Q_1, \dots , Q_4$ of the pencil $\{C_\la\}$ are solutions of both equations
$\theta_0 ({\bf z}) =0, \quad  \theta_1 ({\bf z}) =0$.
 On $\Jac(\Gamma)$ they define two points ${\bf z}={\bf 0},{\bf z}= \frac 12 \tau (0,1)^t$, which give 4 preimages on $\Prym(K,\sigma)$. 

For a particular $\la_*=\mu^*/\nu^*$, equation $ \Xi_{\nu^*, \mu^*} (z)=0$ defines the curve $C\subset \Prym(K,\sigma)$ itself, that is,
\begin{equation} \label{param}
     \Xi_{\nu^*, \mu^*} \left( {\cal A} (P) \right ) \equiv 0 \quad \text{for any $P\in C$.}
\end{equation}
The coefficients $\nu^*, \mu^*$ can be found as follows.
Let $\eps$ be a local parameter near the base point $Q_1=(s_1,0)$ such that $\eps(Q_1)=0$. One can chose $\eps=\sqrt{x-s_1}$. Then, near $Q_1$,
$$
  {\cal A} (P) =\phi(Q_1) \,\eps + O(\eps^2), \quad
\phi(Q_1)= (\phi_1,\phi_2)^t=\frac{d}{d\eps} {\cal A}(P)\Bigl\vert_{P= Q_1}.
$$
In view of the choice of the differentials on $C$, $(\phi_1,\phi_2)^t$ is proportional to $T \cdot (1, s_1)^t$. 
Then, taking derivative of \eqref{param} with respect to $\eps$, we get
$$
\nu^* \left[\frac{\partial \theta_0}{\partial z_1} (0,0) \phi_1   + \frac{\partial  \theta_0}{\partial z_2} (0,0) \phi_2\right]+ \mu^* \left[\frac{\partial \theta_1}{\partial z_1} (0,0) \phi_1   + \frac{\partial  \theta_1}{\partial z_2} (0,0) \phi_2\right]=0,
$$
or
\begin{equation} \label{numu}
\nu^* \,\partial_V \theta_0 (0,0) + \mu^* \, \partial_V \theta_1 (0,0)=0 ,\quad V= T \cdot (1, s_1)^t .
\end{equation}
The vector $V$ and the derivatives
$ \partial_{z_j} \theta_0 (0,0), \partial_{z_j} \theta_1 (0,0)$, $j=1,2$ can be calculated explicitly in terms of theta-functions with the period matrix $\tau$ in \eqref{tau_m}. 

As a result, for $\nu^*,\mu^*$ satisfying \eqref{numu}, the theta-function in \eqref{param} takes the form
\begin{equation} \label{Xi}
\Xi_{\nu^*, \mu^*} ({\bf z})
 = \partial_V \theta_1 ({\bf 0})\, \theta_0 ({\bf z}) -\partial_V \theta_0 ({\bf 0})\, \theta_1 ({\bf z}) .
\end{equation}

In the sequel we will denote this function as just $\Xi^*({\bf z})$.

\paragraph{F. K\"otter's solution in terms of theta-functions.} Using the above notation, the complete solution of the Clebsch system given on page 100 of \cite{kot892} can be written as follows
\begin{gather}
\begin{aligned}
 S_\alpha &= \frac{d_\alpha\, \theta[\eta_\alpha] ({\bf z} + {\cal K}) + e_\alpha\,\theta[\eta_{\alpha\,\beta}] ({\bf z} + {\cal K}) }{\Xi_{\nu^*, \mu^*} ({\bf z})  }\, , \\
T_\alpha &= \frac{\bar d_\alpha\, \theta[\eta_\alpha] ({\bf z} + {\cal K}) + \bar e_\alpha\, \theta[\eta_{\alpha\,\beta}] ({\bf z} + {\cal K}) }{\Xi_{\nu^*, \mu^*} ({\bf z})  } \,, \qquad \alpha=1,2,3, 
\end{aligned} \label{ST_theta}, \\
{\bf z}=T\, \begin{pmatrix}t_s \\ t_p \end{pmatrix}+ {\bf z}_0,
\end{gather}
where $\eta_\alpha, \eta_\beta$ are some of the characteristics from \eqref{deltas}, and $\eta_{\alpha\, \beta}= \eta_\alpha + \eta_\beta \;
\text{mod } \frac 12 {\mathbb Z}^2/{\mathbb Z}^2$.
(We do not give explicit expressions for them and for the constants $d_\alpha, e_\alpha, \bar d_\alpha, e_\alpha$, $\alpha=1,2,3$.) As above, $t_s, t_p$ are the complex times along the flows $X_s, X_p$, and ${\bf z_0}$ is the initial phase.    

Using the quasiperidicity property \eqref{1.5}, one can check that the theta-quotients in
the right hand sides in \eqref{ST_theta} are not meromorphic on $\Prym(K,\sigma)$, but on its 4-fold covering $\Prym(C,\sigma)$,
as stated in Theorem \ref{Th1}. According to the above solution, the variables $S_\alpha, T_\alpha$ have
simple poles along the preimage $\Pi^{-1}(C)$, which is the curve ${\cal D}\subset \Prym(C,\sigma)$ described in the same
theorem.

\paragraph{The quasi-periodic theta-functions on $\Prym(C,\sigma)$.}
By analogy with the theta-function $\Xi_{\nu,\mu}({\bf z})$, introduce the family of functions
\begin{gather}
  \Upsilon_{\nu, \mu} ({\bf z}) = \nu\,  \theta_0 \left(\frac{\bf z}{2}\mid \frac{\tau}{2} \right) +
\mu \,\theta_1\left(\frac{\bf z}{2}\mid \frac{\tau}{2} \right), \\
\theta_0 \left(\frac{\bf z}{2}\mid \frac{\tau}{2} \right)
=\theta \left[\begin{matrix} 0 & 0 \\ 0 & 0 \end{matrix} \right]\left(\frac{\bf z}{2}\mid \frac{\tau}{2} \right), \quad
\theta_1 \left(\frac{\bf z}{2}\mid \frac{\tau}{2} \right)
=\theta \left[\begin{matrix} 0 & 0 \\ 1/2 & 0 \end{matrix} \right]\left(\frac{\bf z}{2}\mid \frac{\tau}{2} \right) ,
 \notag
\end{gather}
$\nu,\mu$ being arbitrary constants.
Here each of the theta functions $\theta_0, \theta_1$ is quasi-periodic on the Jacobian of the curve
$\tilde\Gamma$ in the diagram \eqref{puzzle}, but $\Upsilon_{\nu, \mu} ({\bf z})$ is quasiperiodic only on 2-fold covering of $\Jac(\tilde\Gamma)$, which is
$\Prym(C,\sigma)$\footnote{The argument ${\bf z}/2$ appears due to the rescaling in \eqref{pers_Jacs}.}.

Indeed, in view of \eqref{pers_Jacs} \eqref{periods}, the period matrix of $\Prym(C,\sigma)$ is obtained from that of $\Jac(\tilde\Gamma)$ by doubling the 3rd period vector. Then, in view of \eqref{1.5},
\begin{gather*}
 \theta_0 \left( \frac{\bf z}{2} + \frac{\tau}{2}\,(1,0)^t \right)
= \exp\left(\frac {\tau_{11}}{4 }+z_1\right)\,\theta_0 \left(\frac{\bf z}{2}\right), \quad
\theta_1 \left( \frac{\bf z}{2} + \frac{\tau}{2}\,(1,0)^t \right)
= - \exp \left(\frac {\tau_{11}}{4 }+z_1\right)\,
\theta_1 \left(\frac{\bf z}{2}\right),
\end{gather*}
and when the argument ${\bf z}/2$ changes by any other period,
both theta-functions are multiplied by the same factor.

Hence the zeros of $\Upsilon_{\nu, \mu} ({\bf z})$  on
 $\Prym(C,\sigma)$ are well-defined.
For some particular values $\nu^*,\mu^*$, the equation $\Upsilon_{\nu^*, \mu^*} ({\bf z})=0$
defines the curve $K\subset \Prym(C,\sigma)$, that is
$$
\Upsilon_{\nu^*, \mu^*} ( \bar{\cal A} (R) )\equiv 0 \quad \text{for any $R\in K$}.
$$
Similarly to \eqref{numu}, \eqref{Xi}, one can show that
\begin{equation} \label{theta_Yps}
 \Upsilon_{\nu^*, \mu^*} ({\bf z})
= \partial_v \theta_1 ({\bf 0}| \tau/2)\, \theta_0 \left(\frac{\bf z}{2}\mid \frac{\tau}{2} \right)- \partial_v \theta_0 ({\bf 0}| \tau/2)\, \theta_1 \left(\frac{\bf z}{2}\mid \frac{\tau}{2} \right ) ,
\end{equation}
where $v=T (1,z_1)^t$ is the vector tangent to the base point $Z_1=(0,z_1)$ of the curve $K\subset \Prym(C,\sigma)$.
In the sequel we denote $ \Upsilon_{\nu^*, \mu^*} ({\bf z}) $ as $\Upsilon^* ({\bf z})$.

We note that by using addition formulae for theta-functions with characteristics (see e.g., [Fay84]), the functions
$\theta_0 \left({\bf z}\mid{\tau} \right),
\theta_1\left({\bf z}\mid {\tau}\right)$ can be written as sums of products of
$\theta_0 \left(\frac{\bf z}{2}\mid \frac{\tau}{2} \right),
\theta_1\left(\frac{\bf z}{2}\mid \frac{\tau}{2} \right)$.

\paragraph{Inversion of the Abel--Prym maps $\cal P, \widehat{\cal P}$.}
For $\Xi^*(z)$ introduced in \eqref{Xi}, consider the function
$$
  F(P)=  \Xi^*\left ( e+ {\cal A}(P) \right) , \qquad P\in C,
$$
$e\in {\mathbb C}^2$ being any fixed vector.
$F(P)$ is not single-valued on the curve $C$ as, by its definition, $\Xi^*$ is multiplied by an exponent
when its argument increases by a vector of the lattice $\Lambda$. However, zeros of $F(P)$ on $C$ are well-defined.

\begin{theorem} \label{nullstelle} If $e\ne 0$, the function $F(P)$ has precisely 4 zeros on $C$ (possibly, with multiplicity). \end{theorem}

\noindent{\it Proof.} Let $D\subset A$ be an algebraic curve giving an ample divisor on an Abelian surface $A$ with polarization $(\delta_1, \delta_2)$, and
$L(k D)$ denote the linear space of meromorphic functions on $A$ having poles of degree at most $k$ along $D$ only.

To prove the theorem we use the adjunction formula (see, e.g., \cite{GH}): 
$$
  \text{dim} \, L(D) = \frac 12 D\cdot D =\text{genus} (D)-1 = \delta_1 \delta_2 ,
$$
where $D\cdot D$ is the number of intersections of $D$ with its any translation or with any curve
$D'\subset A$ linearly equivalent to $D$.

Equation $\Xi^*({\bf z})=0$ defines the curve $C\subset \Prym(K,\sigma)$, whereas $e+ {\cal A}(P), \, P\in C$
defines its translation by the vector $e$. Setting in the above formula $D=C$ and $(\delta_1, \delta_2)=(1,2)$,
we see that any two translations of $C$ in $\Prym(K,\sigma)$ intersect at 4 points. $\square$

Next, for a fixed vector $e\in {\mathbb C}^2$ consider the function
$$
  \bar F(R)=  \Xi^*\left ( e+ \Pi \circ \bar {\cal A}(R) \right) =
\text{an exponent}\cdot \Xi^*\left ( e+ \int_{Z_1}^R T \begin{pmatrix} \bar \omega_1 \\ \bar \omega_2 \end{pmatrix}  \right)  , \qquad R\in K,
$$
where, as above, $\Pi$ is the projection $\Prym(C,\sigma)\to\Prym(K,\sigma)$.

\begin{theorem} \label{4R} If $e\ne 0$, $\bar F(R)$
has precisely 8 zeros on the curve $K$ (possibly, with multiplicity). \end{theorem}

\noindent{\it Proof.} Observe that, for $R\in K$, the image $e+ \Pi\circ \bar {\cal A}(R)$ describes a translation
$\hat K_e$ of $\Pi(K)$ in $\Prym(K,\sigma)$.
Then $\bar F(R)$ vanishes exactly at the intersection points $C \cap \hat K_e$.

Note that $C\cdot \Pi(K)= \Pi^{-1}(C')\cdot K'$ for any curves $C', K'$ linearly equivalent to $C$, respectively $K$.
Next, the pullback $\Pi^{-1}(C')$ is linearly equivalent to $2K\subset \Prym(C,\sigma)$. 

Hence, $\Pi^{-1}(C')\cdot K' = 2K \cdot K =2 (K\cdot K)$, which, by the adjunction formula, equals $2\cdot 4=8$.
$\square$

\medskip
Now return to the Abel--Prym map \eqref{APM} and observe that it can be written in the form
\begin{equation} \label{z2}
{\cal P}\, :\; (P\in C, R\in K) \mapsto
{\bf z} =(z_1, z_2)^t= {\cal A}(P)+ \Pi\circ \bar {\cal A}(R) .
\end{equation}

\begin{theorem} \label{main} Under the map $\cal P$, any ${\bf z} \in \Prym(K,\sigma)$ has precisely 8 preimages \\
$(P_1, R_1), \dots, (P_8, R_8)$ on $C\times K$ (possibly, with multiplicity)\footnote{One can compare this with the standard
Abel--Jacobi map $G^{(g)} \to \Jac(G)$, where $G$ is a smooth genus $g$ curve and $G^{(g)}$ is its symmetric power.
A generic point in $\Jac(G)$ has just one pre-image on $G^{(g)}$.}.
The points $R_1,\dots, R_8$ are zeros of the function
$$
{\cal F} (R) =
\Xi^* \left ( {\bf z} - \bar{\cal A}(R) \right), \qquad R\in K.
$$
For each $R_i$, the corresponding point $P_i\in C$ is found as the unique solution of the transcendental equation
\begin{equation} \label{tra}
  {\cal A}(p) \equiv \int_{Q_1}^p  T \begin{pmatrix} \omega_1 \\ \omega_2 \end{pmatrix} = {\bf z} - \bar {\cal A}(R_i),
\qquad p\in C
\end{equation}
\end{theorem}

\paragraph{Remark.} In practice, Theorem \ref{main} allows to calculate only symmetric functions of coordinates of $R_1,\dots, R_8$, i.e., functions which are invariant with respect to permutations of the points. Such functions can be expressed in terms of the theta-function $\Xi^*( {\bf z})$ and its derivatives, in the same way as for the case of inversion of the standard Abel map, see e.g., \cite{Dub}).
\medskip

\noindent{\it Proof of Theorem} \ref{main}.
In view of Theorem \ref{4R}, for any ${\bf z}\in {\mathbb C}^2$ the function ${\cal F} (Q)$ on $K$ has 8 zeros,
denoted as $R_1,\dots, R_8$. Then, for each $R_i$ the vector ${\bf z} - \bar {\cal A}(R_i)$ belongs to the
zero locus of $\Xi^*$, that is, following \eqref{param}, to $C\subset \Prym(K,\sigma)$. Since the map
$\cal A$ is injective, there is a unique solution to \eqref{tra} giving $P_i\in C$ such that
${\cal A}(P_i)+ \Pi\circ \bar {\cal A}(R_i)={\bf z}$. $\square$
\medskip

Although the map $\cal P$ is not one-to-one, it also has an obvious ''injectivity'' property: for any ${\bf z}$ and $\tilde{\bf z}\in \Prym(K,\sigma)$
\begin{equation} \label{injective}
 {\bf z} \ne \tilde{\bf z} \; \text{mod} \;\Lambda \;
\Longrightarrow \; \{ {\cal P}^{-1}({\bf z}) \} \cap
\{ {\cal P}^{-1}(\tilde{\bf z}) \} = \emptyset .
\end{equation}
Indeed, assume that a pair $(P^*, R^*)\in C\times K$ belongs to the intersection
$\{ {\cal P}^{-1}({\bf z}) \} \cap
\{ {\cal P}^{-1}(\tilde{\bf z}) \}$. This means that
${\cal P} (P^*, R^*)$ gives both $\bf z$ and $\tilde{\bf z}$,
which is not possible, because ${\cal P}$ is well-defined.

\paragraph{Inversion of the extended Abel--Prym map.}
Theorem \ref{main} itself does not allow to express symmetric functions of coordinates of $P_1,\dots, P_8\in C$ in terms
of theta-functions. 
This becomes possible by considering inversion of
the map $\widehat{\cal P} \, : \; {\cal D} \times K \to \Prym(C,\sigma)$.
Namely, the following analog of Theorem \ref{main} holds.

\begin{theorem}\label{ex_main} Under the map $\widehat{\cal P}$, any ${\bf z} \in \Prym(C,\sigma)$
has precisely 8 preimages \\
$(P_1^*, R_1), \dots, (P_8^*, R_8)$ on ${\cal D}\times K$ (possibly, with multiplicity).
The points $P_1^*,\dots, P_8^*\in {\cal D}$ are zeros of the function
$$
{\cal F} (P) =
\Upsilon^* \left ( {\bf z} - \widehat{\cal A}(P) \right), \qquad P\in {\cal D},
$$
with $\Upsilon^*( {\bf z})$ defined in \eqref{theta_Yps}.

For each $P_i^*$, the corresponding point $R_i\in K$ is found as the unique solution of the transcendental equation
\begin{equation} \label{tra1}
  \bar {\cal A}(q) \equiv \int_{Z_1}^q  T \begin{pmatrix} \bar \omega_1 \\ \bar \omega_2 \end{pmatrix}
  = {\bf z} - \widehat{\cal A}(P_i^*), \qquad q\in K .
\end{equation}
\end{theorem}

Proof of this theorem goes along the same lines as that of Theorem \ref{main} describing the inversion of the map $\cal P$. Again, in practice, Theorem \ref{ex_main} allows to calculate only symmetric functions of the coordinates of
$P_1^*,\dots, P_8^*\in {\cal D}$ in terms of $\Upsilon^*( {\bf z})$ and its derivatives.

Similarly to $\cal P$, the extended map $\hat{\cal P}$ enjoys the injectivity property
\begin{equation} \label{injective1}
 {\bf z} \ne \tilde{\bf z} \; \text{mod} \,\bar\Lambda \;
\Longrightarrow \{ \hat{\cal P}^{-1}({\bf z}) \} \cap
\{ \hat{\cal P}^{-1}(\tilde{\bf z}) \} = \emptyset .
\end{equation}

Both theorems, together with the relation \eqref{comm}, lead to following property.

\begin{proposition} \label{comD} For any ${\bf z} \in \Prym(C,\sigma)$, let $\widehat {\bf z}= \Pi({\bf z})\in \Prym (K,\sigma)$. Then, if
$ \widehat{\cal P}^{-1} ({\bf z}) =\{(P_1^*, R_1), \dots, (P_8^*, R_8)\}$, one has
${\cal P}^{-1}(\widehat {\bf z}) =\{(P_1, R_1), \dots, (P_8, R_8)\}$ such that $P_i=\Pi(P_i^*), \,i=1,\dots,8$. That is, the following diagram holds
$$
\begin{CD}
 \{(P^*_k,R_k), \, k=1,\dots,8 \} @ < \widehat{\cal P}^{-1} <<  {\bf z} \in  \Prym(C,\sigma)  \\
       @  V \Pi VV  @  V  \Pi^* VV  \\
  \{(P_k,R_k), \, k=1,\dots,8 \} @ < {\cal P}^{-1}  <<  \widehat{\bf z} \in  \Prym(K,\sigma)  
 \end{CD}
$$
\end{proposition}

\section{Theta-function solution via the separation of variables.} As stated in Theorem \ref{Th1}, there is a bijection between the compactified complex invariant torus ${\cal I}_H \cup {\cal D}$ and $\Prym(C,\sigma)$, and between the compactified factor variety $({\cal I}_H/{\mathfrak g})\cup C$ and $\Prym(K,\sigma)$. On the other hand, as we recalled in Introduction, each point on ${\cal I}_H$ gives rise to 8 sets $\{(x_1,w), (x_2,W)\}$ of the separating variables constructed in Section 4 of Part I, i.e., there is a correspondence
$$
 {\cal S}\, : \;  (S_\alpha^2, T_\alpha^2) \to \text{8 sets $\{(x_1,w), (x_2,W)\}$} \,. 
$$ 
Its inversion is described by the reconstruction formulas (38, Part I). 

It is then natural to expect that for any ${\bf z}\in \Prym(K,\sigma)$, the coordinates of the eight preimages on $C\times K$,
$\{(P_1, R_1), \dots, (P_8, R_8)\} ={\cal P}^{-1}({\bf z})$
give precisely the above eight sets. Indeed, the following theorem holds. 

\begin{theorem} \label{bi1} 1) The coordinates $(x_1,w), (x_2,W)$ of any of the preimages \\ $(P_1, R_1), \dots, (P_8, R_8)$ of ${\bf z} \in \Prym(K,\sigma)$ give the same values of the squares $S_\alpha^2, T_\alpha^2$, $\alpha=1,2,3$ via the reconstruction formulas (38, Part I).

Equivalently, under the map $\cal P$, the 8 sets of the separation variables  $(x_1,w)\in C, (x_2,W)\in K$, corresponding to a point on ${\cal I}_H$, give the same point ${\bf z} \in \Prym(K,\sigma)$.

2) The coordinates $(x_1,w,v_1,v_2,v_3), (x_2,W)$ of any of the preimages $(P_1^*, R_1)$, $\dots$, $(P_8^*, R_8) \in {\cal D}\times K$ of
${\bf z}^* \in \Prym(C,\sigma)$ give the same values of $S_\alpha, T_\alpha$, $\alpha=1,2,3$ via the same reconstruction formulas.
\end{theorem}

\noindent{\it Proof.} For any fixed point $(\bar S_\alpha^2, \bar T_\alpha^2)\in ({\cal I}_H/{\mathfrak g})\cup C$, let 
$\{(x_1^*,w^*), (x_2^*,W^*)\}$ and \\ $\{(x_1^{**},w^{**}), (x_2^{**},W^{**})\}$ be any two sets of ${\cal S}(\bar S_\alpha^2, \bar T_\alpha^2)$. Let now 
$$
{\bf z}^* = {\cal P} ((x_1^*,w^*), (x_2^*,W^*)), \quad 
{\bf z}^{**} ={\cal P} ((x_1^{**},w^{**}), (x_2^{**},W^{**})). 
$$   
Assume that ${\bf z}^* \ne {\bf z}^{**}$ (mod $\Lambda$). Then, by the injectivity property \eqref{injective}, \\
$
 \{ {\cal P}^{-1}({\bf z}^*) \} \cap
\{ {\cal P}^{-1}({\bf z}^{**}) \} = \emptyset$, {hence}
$$
\{(x_1^{**},w^{**}), (x_2^{**},W^{**})\} \ne \{ {\cal P}^{-1}({\bf z}^*) \} , \quad \{(x_1^{*},w^{*}), (x_2^{*},W^{*})\} \ne \{ {\cal P}^{-1}({\bf z}^{**}) \}  .
 $$
It follows that the reconstruction ${\cal S}^{-1}$ applied to $\{ {\cal P}^{-1}({\bf z}^*) \}$ gives different values $(S_\alpha^2, T_\alpha^2)$, because otherwise, together with $\{(x_1^{**},w^{**}), (x_2^{**},W^{**})\}$, there were 9 different sets giving the same $(\bar S_\alpha^2, \bar T_\alpha^2)$. But this contradicts the fact that the composition of $\cal S$ and the map $\cal P$ realizes a bijection between $({\cal I}_H/{\mathfrak g})\cup C$ and $\Prym(K,\sigma)$. Therefore our assumption  ${\bf z}^* \ne {\bf z}^{**}$ was wrong. This proves item 1).

Item 2) follows from the the same arguments and Proposition \ref{comD}. $\square$       
\medskip

Theorems \ref{bi1} and \ref{ex_main} enables one to obtain theta-function solution to the Clebsch system in terms of theta-functions. To do this we only need 
to express the variables $S_\alpha, T_\alpha$ in terms of symmetric functions of coordinates of the points $P_1^*, \dots, P_8^*$ on $\cal D$, the preimages of ${\bf z} \in \Prym(C,\sigma)$. According to Theorem \ref{ex_main}, the latter functions can be written in terms of $\Upsilon^*( {\bf z})$ in \eqref{theta_Yps} and its derivatives.

Note however, that the structure of the obtained theta-function solutions will be 
different from K\"otter's solutions \eqref{ST_theta}, as the latter are written in terms of $\Upsilon^*(z)$. 

Our main tool will be again the constraint equation (22, Part I), defining the submanifold $\hat{\cal S}$ in the extended phase space $\hat M$:   
\begin{gather} \label{vinc}
{\cal F}(x)= \sum_{\alpha=1}^3 c_\alpha ( v_\alpha S_\alpha + v_\beta v_\gamma T_\alpha) =0 , \quad
v_\alpha^2 = x+j_\alpha, \quad 
c_\alpha^2=\frac{1}{(j_\alpha-j_\beta)(j_\alpha-j_\gamma)}. 
\end{gather} 
Here, as above, the radicals $v_\alpha=\sqrt{x+j_\alpha}$, $\alpha=1,2,3$ are meromorphic functions on the elliptic curve $\hat E$. 
Substituting their parameterization \eqref{vi} into \eqref{vinc}, we obtain 
\begin{gather}
{\cal F}=(f_{12} Z^2+ f_{11}Z+f_{10}) Y 
+ f_{04}Z^4+ f_{03}Z^3+f_{02}Z^2 +f_{01} Z+f_{00}=0\, , \label{calF} 
\end{gather}
with
\begin{gather}
f_{12}= 2(\hat S_1+\hat S_2+\hat S_3), \; f_{11}=4(j_1\hat S_1+j_2 \hat S_2+j_3\hat S_3), \;   
 f_{10}= 2\sum_{ (\alpha,\beta,\gamma) } j_\alpha j_\beta (\hat S_\alpha+\hat S_\beta-\hat S_\gamma), \notag \\
f_{04}= \hat T_1+\hat T_2+\hat T_3, \quad 
f_{03}= 2 \sum_{(\alpha,\beta,\gamma)} (j_\alpha+j_\beta)\hat T_\gamma,\quad 
f_{02}= 6 \sum_{(\alpha,\beta,\gamma) } j_\alpha j_\beta\,\hat T_\gamma, \label{f0123} \\ 
f_{01}=2 \sum_{(\alpha,\beta,\gamma)}
[ j_\gamma j_\beta (j_1+j_2+j_3)+ j_1 j_2 j_3
-j_\alpha(j_\gamma^2+j_\beta^2)]\, \hat T_\alpha , \notag \\ 
f_{00}=\sum_{(\alpha,\beta,\gamma)} ( j_\alpha j_\gamma+j_\beta j_\gamma-j_\alpha j_\beta)(j_\alpha j_\beta+j_\beta j_\gamma-j_\alpha j_\gamma ) \hat T_\alpha =0, \notag
\end{gather}
where we set
$  \hat S_\alpha= c_\alpha\, S_\alpha, \;
\hat T_\alpha= c_\alpha\, T_\alpha$,
$ (\alpha,\beta,\gamma)=(1,2,3)$.
Since the variables $Z,Y$ are constrained by the equation of $\hat E$, for each generic $T,S$ the equation \eqref{calF} has precisely 8 solutions $(Z_1,Y_1),\dots,(Z_8,Y_8)$, which are projections of the preimages $P_1^*,\dots, P_8^*\in {\cal D}$ onto $\widehat E$ (see the equation \eqref{D2} of ${\cal D}$).   

Substituting these solutions into \eqref{calF} subsequently, one obtains a system of 8 linear homogeneous equations for the eight coefficients $f_{12},\dots,f_{00}$. By construction, the rank of the system is 6.  
Solving it and inverting the linear relations \eqref{f0123},  one finds the variables $\hat S_\alpha, \hat T_\alpha$ and, therefore, $S_\alpha, T_\alpha$, as symmetric functions of $(Z_1,Y_1),\dots,(Z_8,Y_8)$ up to a common factor $\varkappa$: 
\begin{equation} \label{symm_ST}
 S_\alpha = \frac{\Sigma_\alpha }{\varkappa}, \quad
T_\alpha = \frac{\Sigma_{\alpha+3} }{\varkappa}, \qquad
\alpha=1,2,3 .
\end{equation}  
Explicit expressions for the symmetric functions $\Sigma_1,\dots, \Sigma_6$ are quite long, so we do not give them here. 

Next, we substitute \eqref{symm_ST} into the area integral $S_1 T_1+ S_2 T_2+ S_3 T_3 = C_1$ and obtain $\varkappa^2$ as a symmetric function of $(Z_1,Y_1),\dots,(Z_8,Y_8)$ as well.
Finally, by using equation \eqref{D2} of the curve ${\cal D}\subset {\mathbb C}^3(Z,Y,G)$, this function can be shown to be a full square of a symmetric function of $(Z_1,Y_1,G_1),\dots,(Z_8,Y_8,G_8)$, the coordinates of the points $P_1^*,\dots, P_8^*\in {\cal D}$. 
This, together with \eqref{symm_ST}, provides us with the expresions for $S_\alpha, T_\alpha$ we needed. 

The latter can be regarded as a (much more cumbersome) alternative to the reconstruction formulas (38, Part I). There is a more essential difference although: whereas these formulas give the original variables $S_\alpha, T_\alpha$  in terms of just one pair of points on both curves, ${\cal D}$ and $K$, the expressions \eqref{symm_ST} involve 8 points, but on ${\cal D}$ only.    

\subsection*{Acknowledgments} Y. Fedorov is grateful to 
V. Enolski, A. Izosimov, T. Shaska for stimulating discussions.   His contribution was partially supported by the Spanish MINECO-FEDER Grant MTM2015-65715-P and the Catalan grant 2017SGR1049. 
The authors also acknowledge the support and hospitality of the Department of Mathematics of Universit\'{a} di Milano Bicocca, where a part of this work had been made.


\begin{thebibliography}{30}

\bibitem[AvM984]{avm84}  Adler M., van Moerbeke P.: Geodesic flow on $so(4)$ and the intersections
of quadrics. {\it Proc.Natl. Acad. Sci. USA.\/} {\bf 81}, (1984), 4613--4616

\bibitem[Aud996]{au96}
Audin, M. {\it Spinning tops. A course on integrable systems.} Cambridge Studies in Advanced Mathematics, 51.
Cambridge University Press, Cambridge, 1996

\bibitem[Bar985]{bw85} Barth, W. Abelian surfaces with $(1,2)$-polarization. Algebraic geometry, Sendai, 1985, 41--84, Adv. Stud. Pure Math.,
{\bf 10}, North-Holland, Amsterdam, 1987.

\bibitem[BBEIM]{Bel} Belokolos E.D., Bobenko A.I., Enol'sii V.Z., Its A.R., and   Matveev V.B.
{\it Algebro-Geometric Approach to Nonlinear   Integrable Equations.\/}
 Springer Series in Nonlinear Dynamics.   Springer--Verlag 1994.

\bibitem[Bob86]{Bob_so4}  Bobenko A.I. Euler equations in the Lie algebras $e(3)$ and $so(4)$.  
Isomorphisms of integrable cases.{\it Funkts. Anal. Prolozh.\/}  {\bf 20}, No.1,\,
64-66 (1986). English transl.:{\it Funct. Anal. Appl.\/}  {\bf 20} (1986), 53-56\,

\bibitem[Dub81]{Dub} Dubrovin B.A. Theta-functions and non-linear equations. {\it Usp.Mat. Nauk\/} 
{\bf 36}, No.2 (1981), 11-80. English transl.: {\it Russ. Math. Surveys\/}. {\bf 36} (1981), 11-92
 

\bibitem[EF17]{ef17} Enolski V., Fedorov Yu.
Algebraic description of Jacobians isogeneous to certain Prym varieties with polarization (1,2).
{\it Exp. Math.} {\bf 27} (2018), no. 2, 147--178

\bibitem[Fay84]{fay} Fay  J.  Theta-functions on Riemann Surfaces.
{\it Springer Lecture Notes\/} {\bf 352}, Springer-Verlag, 1973

\bibitem[Fed95]{Fe_AMS}  Fedorov Yu. Integrable systems, Lax representations, and confocal
   quadrics. In: {\it Dynamical systems in classical mechanics\/}, 173--199,
   Amer. Math. Soc. Transl. Ser. 2, {\bf 168}, Amer. Math. Soc., Providence, RI,   1995

\bibitem[GH]{GH} Griffits, P., Harris, J. 
{\it Principles of Algebraic Geometry.\/} 
Wiley Interscience, New York 1978

\bibitem[Hai983]{hai83} Haine L. Geodesic flow on $so(4)$ and Abelian surfaces.
{\it Math. Ann.} \textbf{263} (1983), 435--472.

\bibitem[HvM989]{hvm89} Horozov E., van Moerbeke P.
The full geometry of Kowalewski's top and $(1,2)$-abelian surfaces.
{\it Comm. Pure Appl. Math.} \textbf{42}:4 (1989) 357--407.


\bibitem[Mum974]{Mum} Mumford D. Prym Varieties I. in: {\it Contributions to analysis}, 
Ahlfors L.V. Kra I. Maskit B. Nirenberg
L., Eds., Academic Press (1974), 325--350.

\bibitem[Mum984]{MumTh} Mumford D. Tata Lectures on Theta II. {\it Progress in Math.\/}{\bf 43}, 1984 

\bibitem[Kot892]{kot892} F.~K\"otter. Uber die Bewegung eines festen K\"orpers in einer
Fl\"ussigkeit. I, II. {\it J. reine angew. Math.} {\bf 109} (1892), 51--81, 89--111.

\bibitem[Pan986]{Pant} Pantazis S. Prym varieties and the geodesic flow on $SO(n)$.  {\it Math. Ann.} (1986) 273--297.

\bibitem[Per81]{Per}  Perelomov A. M. Some remarks on the integrability of the equations
 of motion of a rigid body in an ideal liquid. {\it Funkts.Anal.i Prilozh.\/}
{\bf 15},  No.2 (1981), 83-85. English transl.: Funct.Anal.Appl. 15, No.2 \/
 (1981), 83-85





\bibitem[Web878]{web878} Weber H.  Anwendung der Thetafunctionen zweir Veranderlicher auf
die Theorie der Bewegung eines festen K\"orpers in einer Fl\"ussigkeit.
{\it Math. Ann.} \textbf{14} (1878), 173--206

\end{thebibliography}
\end{document}